\documentclass[nofootinbib,prd,twocolumn]{revtex4}%
\usepackage{amsmath}
\usepackage{amsfonts}
\usepackage{amssymb}
\usepackage{graphicx}%
\begin{document}
\title{Casimir Energy and the Cosmological Constant}
\author{Remo Garattini}
\email{Remo.Garattini@unibg.it}
\affiliation{Universit\`{a} degli Studi di Bergamo, Facolt\`{a} di Ingegneria, Viale
Marconi 5, 24044 Dalmine (Bergamo) ITALY.}

\begin{abstract}
We regard the Wheeler-De Witt equation as a Sturm-Liouville problem with the
cosmological constant considered as the associated eigenvalue. The used method
to study such a problem is a variational approach with Gaussian trial wave
functionals. We approximate the equation to one loop in a Schwarzschild
background. A zeta function regularization is involved to handle with
divergences. The regularization is closely related to the subtraction
procedure appearing in the computation of Casimir energy in a curved
background. A renormalization procedure is introduced to remove the infinities
together with a renormalization group equation.

\end{abstract}
\maketitle

\section{Introduction}

One of the most fascinating and unsolved problems of the theoretical physics
of our century is the cosmological constant. Einstein introduced his
cosmological constant $\Lambda_{c}$ in an attempt to generalize his original
field equations. The modified field equations are
\begin{equation}
R_{\mu\nu}-\frac{1}{2}g_{\mu\nu}R+\Lambda_{c}g_{\mu\nu}=8\pi GT_{\mu\nu},
\label{i0}%
\end{equation}
where $\Lambda_{c}$ is the cosmological constant, $G$ is the gravitational
constant and $T_{\mu\nu}$ is the energy-momentum tensor. By redefining
\begin{equation}
T_{\mu\nu}^{tot}\equiv T_{\mu\nu}-\frac{\Lambda_{c}}{8\pi G}g_{\mu\nu},
\end{equation}
one can regain the original form of the field equations
\begin{equation}
R_{\mu\nu}-\frac{1}{2}g_{\mu\nu}R=8\pi GT_{\mu\nu}^{tot}=8\pi G\left(
T_{\mu\nu}+T_{\mu\nu}^{\Lambda}\right)  , \label{i2}%
\end{equation}
at the prize of introducing a vacuum energy density and vacuum stress-energy
tensor
\begin{equation}
\rho_{\Lambda}=\frac{\Lambda_{c}}{8\pi G};\qquad T_{\mu\nu}^{\Lambda}%
=-\rho_{\Lambda}g_{\mu\nu}.
\end{equation}
Alternatively, Eq.$\left(  \ref{i0}\right)  $ can be cast into the form,%
\begin{equation}
R_{\mu\nu}-\frac{1}{2}g_{\mu\nu}R+\Lambda_{eff}g_{\mu\nu}=0,
\end{equation}
where we have included the contribution of the vacuum energy density in the
form $T_{\mu\nu}=-\left\langle \rho\right\rangle g_{\mu\nu}$. In this case
$\Lambda_{c}$ can be considered as the bare cosmological constant
\begin{equation}
\Lambda_{eff}=8\pi G\rho_{eff}=\Lambda_{c}+8\pi G\left\langle \rho
\right\rangle .
\end{equation}
Experimentally, we know that the effective energy density of the universe
$\rho_{eff}$ is of the order $10^{-47}GeV^{4}$. A crude estimate of the Zero
Point Energy (ZPE) of some field of mass $m$ with a cutoff at the Planck scale
gives%
\begin{equation}
E_{ZPE}=\frac{1}{2}\int_{0}^{\Lambda_{p}}\frac{d^{3}k}{\left(  2\pi\right)
^{3}}\sqrt{k^{2}+m^{2}}\simeq\frac{\Lambda_{p}^{4}}{16\pi^{2}}\approx
10^{71}GeV^{4} \label{zpe}%
\end{equation}
This gives a difference of about 118 orders\cite{Lambda}. The approach to
quantization of general relativity based on the following set of equations%
\begin{equation}
\left[  \frac{2\kappa}{\sqrt{g}}G_{ijkl}\pi^{ij}\pi^{kl}-\frac{\sqrt{g}%
}{2\kappa}\left(  R-2\Lambda_{c}\right)  \right]  \Psi\left[  g_{ij}\right]
=0 \label{WDW}%
\end{equation}
and%
\begin{equation}
-2\nabla_{i}\pi^{ij}\Psi\left[  g_{ij}\right]  =0, \label{diff}%
\end{equation}
where $R$ is the three-scalar curvature, $\Lambda_{c}$ is the bare
cosmological constant and $\kappa=8\pi G$, is known as Wheeler-De Witt
equation (WDW)\cite{De Witt}. Eqs.$\left(  \ref{WDW}\right)  $ and $\left(
\ref{diff}\right)  $ describe the \textit{wave function of the universe}. The
WDW equation represents invariance under \textit{time} reparametrization in an
operatorial form, while Eq.$\left(  \ref{diff}\right)  $ represents invariance
under diffeomorphism. $G_{ijkl}$ is the \textit{supermetric} defined as%
\begin{equation}
G_{ijkl}=\frac{1}{2}(g_{ik}g_{jl}+g_{il}g_{jk}-g_{ij}g_{kl}).
\end{equation}
Note that the WDW equation can be cast into the form%
\begin{equation}
\left[  \frac{2\kappa}{\sqrt{g}}G_{ijkl}\pi^{ij}\pi^{kl}-\frac{\sqrt{g}%
}{2\kappa}R\right]  \Psi\left[  g_{ij}\right]  =-\frac{\sqrt{g}}{\kappa
}\Lambda_{c}\Psi\left[  g_{ij}\right]  ,
\end{equation}
which formally looks like an eigenvalue equation. In this paper, we would like
to use the Wheeler-De Witt (WDW) equation to estimate $\left\langle
\rho\right\rangle $. In particular, we will compute the gravitons ZPE
propagating on the Schwarzschild background. This choice is dictated by
considering that the Schwarzschild solution represents the only non-trivial
static spherical symmetric solution of the Vacuum Einstein equations.
Therefore, in this context the ZPE can be attributed only to quantum
fluctuations. The used method will be a variational approach applied on
gaussian wave functional. The rest of the paper is structured as follows, in
section \ref{p1}, we show how to apply the variational approach to the
Wheeler-De Witt equation and we give some of the basic rules to solve such an
equation approximated to second order in metric perturbation, in section
\ref{p2}, we analyze the spin-2 operator or the operator acting on transverse
traceless tensors, in section \ref{p3}, we analyze the spin-0 operator, in
section \ref{p4} we use the zeta function to regularize the divergences coming
from the evaluation of the ZPE for TT tensors and we write the renormalization
group equation, in section \ref{p5} we use the same procedure of section
\ref{p4}, but for the trace part. We summarize and conclude in section
\ref{p6}.

\section{The Wheeler-De Witt equation and the cosmological constant}

\label{p1}The WDW equation $\left(  \ref{WDW}\right)  $, written as an
eigenvalue equation, can be cast into the form%
\begin{equation}
\hat{\Lambda}_{\Sigma}\Psi\left[  g_{ij}\right]  =\Lambda^{\prime}\Psi\left[
g_{ij}\right]  , \label{WDW1}%
\end{equation}
where
\begin{equation}
\left\{
\begin{array}
[c]{c}%
\hat{\Lambda}_{\Sigma}=\frac{2\kappa}{\sqrt{g}}G_{ijkl}\pi^{ij}\pi^{kl}%
-\frac{\sqrt{g}}{2\kappa}R\\
\\
\Lambda^{\prime}=-\frac{\Lambda}{\kappa}%
\end{array}
\right.  .
\end{equation}
We, now multiply Eq.$\left(  \ref{WDW1}\right)  $ by $\Psi^{\ast}\left[
g_{ij}\right]  $ and we functionally integrate over the three spatial metric
$g_{ij}$, then after an integration over the hypersurface $\Sigma$, one can
formally re-write the WDW equation as%
\begin{equation}
\frac{1}{V}\frac{\int\mathcal{D}\left[  g_{ij}\right]  \Psi^{\ast}\left[
g_{ij}\right]  \int_{\Sigma}d^{3}x\hat{\Lambda}_{\Sigma}\Psi\left[
g_{ij}\right]  }{\int\mathcal{D}\left[  g_{ij}\right]  \Psi^{\ast}\left[
g_{ij}\right]  \Psi\left[  g_{ij}\right]  }=\frac{1}{V}\frac{\left\langle
\Psi\left\vert \int_{\Sigma}d^{3}x\hat{\Lambda}_{\Sigma}\right\vert
\Psi\right\rangle }{\left\langle \Psi|\Psi\right\rangle }=\Lambda^{\prime}.
\label{WDW2}%
\end{equation}
The formal eigenvalue equation is a simple manipulation of Eq.$\left(
\ref{WDW}\right)  $. However, we gain more information if we consider a
separation of the spatial part of the metric into a background term, $\bar
{g}_{ij}$, and a perturbation, $h_{ij}$,%
\begin{equation}
g_{ij}=\bar{g}_{ij}+h_{ij}.
\end{equation}
Thus eq.$\left(  \ref{WDW2}\right)  $ becomes%
\begin{equation}
\frac{\left\langle \Psi\left\vert \int_{\Sigma}d^{3}x\left[  \hat{\Lambda
}_{\Sigma}^{\left(  0\right)  }+\hat{\Lambda}_{\Sigma}^{\left(  1\right)
}+\hat{\Lambda}_{\Sigma}^{\left(  2\right)  }+\ldots\right]  \right\vert
\Psi\right\rangle }{\left\langle \Psi|\Psi\right\rangle }=\Lambda^{\prime}%
\Psi\left[  g_{ij}\right]  , \label{WDW3}%
\end{equation}
where $\hat{\Lambda}_{\Sigma}^{\left(  i\right)  }$ represents the $i^{th}$
order of perturbation in $h_{ij}$. By observing that the kinetic part of
$\hat{\Lambda}_{\Sigma}$ is quadratic in the momenta, we only need to expand
the three-scalar curvature $\int d^{3}x\sqrt{g}R^{\left(  3\right)  }$ up to
quadratic order and we get%
\[
\int_{\Sigma}d^{3}x\sqrt{\bar{g}}\left[  -\frac{1}{4}h\triangle h+\frac{1}%
{4}h^{li}\triangle h_{li}-\frac{1}{2}h^{ij}\nabla_{l}\nabla_{i}h_{j}%
^{l}\right.
\]%
\begin{equation}
\left.  +\frac{1}{2}h\nabla_{l}\nabla_{i}h^{li}-\frac{1}{2}h^{ij}R_{ia}%
h_{j}^{a}+\frac{1}{2}hR_{ij}h^{ij}+\frac{1}{4}h\left(  R^{\left(  0\right)
}\right)  h\right]  \label{rexp}%
\end{equation}
where $h$ is the trace of $h_{ij}$ and $R^{\left(  0\right)  }$ is the three
dimensional scalar curvature. To explicitly make calculations, we need an
orthogonal decomposition for both $\pi_{ij\text{ }}$and $h_{ij}$ to
disentangle gauge modes from physical deformations. We define the inner product%

\begin{equation}
\left\langle h,k\right\rangle :=\int_{\Sigma}\sqrt{g}G^{ijkl}h_{ij}\left(
x\right)  k_{kl}\left(  x\right)  d^{3}x,
\end{equation}
by means of the inverse WDW metric $G_{ijkl}$, to have a metric on the space
of deformations, i.e. a quadratic form on the tangent space at $h_{ij}$, with%

\begin{equation}%
\begin{array}
[c]{c}%
G^{ijkl}=(g^{ik}g^{jl}+g^{il}g^{jk}-2g^{ij}g^{kl})\text{.}%
\end{array}
\end{equation}
The inverse metric is defined on cotangent space and it assumes the form%

\begin{equation}
\left\langle p,q\right\rangle :=\int_{\Sigma}\sqrt{g}G_{ijkl}p^{ij}\left(
x\right)  q^{kl}\left(  x\right)  d^{3}x,
\end{equation}
so that%

\begin{equation}
G^{ijnm}G_{nmkl}=\frac{1}{2}\left(  \delta_{k}^{i}\delta_{l}^{j}+\delta
_{l}^{i}\delta_{k}^{j}\right)  .
\end{equation}
Note that in this scheme the \textquotedblleft inverse
metric\textquotedblright\ is actually the WDW metric defined on phase space.
The desired decomposition on the tangent space of 3-metric
deformations\cite{BergerEbin,York,MazurMottola,Vassilevich} is:%

\begin{equation}
h_{ij}=\frac{1}{3}hg_{ij}+\left(  L\xi\right)  _{ij}+h_{ij}^{\bot}
\label{p21a}%
\end{equation}
where the operator $L$ maps $\xi_{i}$ into symmetric tracefree tensors%

\begin{equation}
\left(  L\xi\right)  _{ij}=\nabla_{i}\xi_{j}+\nabla_{j}\xi_{i}-\frac{2}%
{3}g_{ij}\left(  \nabla\cdot\xi\right)  .
\end{equation}
Thus the inner product between three-geometries becomes
\[
\left\langle h,h\right\rangle :=\int_{\Sigma}\sqrt{g}G^{ijkl}h_{ij}\left(
x\right)  h_{kl}\left(  x\right)  d^{3}x=
\]%
\begin{equation}
\int_{\Sigma}\sqrt{g}\left[  -\frac{2}{3}h^{2}+\left(  L\xi\right)
^{ij}\left(  L\xi\right)  _{ij}+h^{ij\bot}h_{ij}^{\bot}\right]  . \label{p21b}%
\end{equation}
With the orthogonal decomposition in hand we can define the trial wave
functional as%
\begin{equation}
\Psi\left[  h_{ij}\left(  \overrightarrow{x}\right)  \right]  =\mathcal{N}%
\Psi\left[  h_{ij}^{\bot}\left(  \overrightarrow{x}\right)  \right]
\Psi\left[  h_{ij}^{\Vert}\left(  \overrightarrow{x}\right)  \right]
\Psi\left[  h_{ij}^{trace}\left(  \overrightarrow{x}\right)  \right]  ,
\label{twf}%
\end{equation}
where
\begin{equation}%
\begin{array}
[c]{c}%
\Psi\left[  h_{ij}^{\bot}\left(  \overrightarrow{x}\right)  \right]
=\exp\left\{  -\frac{1}{4}\left\langle hK^{-1}h\right\rangle _{x,y}^{\bot
}\right\} \\
\\
\Psi\left[  h_{ij}^{\Vert}\left(  \overrightarrow{x}\right)  \right]
=\exp\left\{  -\frac{1}{4}\left\langle \left(  L\xi\right)  K^{-1}\left(
L\xi\right)  \right\rangle _{x,y}^{\Vert}\right\} \\
\\
\Psi\left[  h_{ij}^{trace}\left(  \overrightarrow{x}\right)  \right]
=\exp\left\{  -\frac{1}{4}\left\langle hK^{-1}h\right\rangle _{x,y}%
^{Trace}\right\}
\end{array}
.
\end{equation}
The symbol \textquotedblleft$\perp$\textquotedblright\ denotes the
transverse-traceless tensor (TT) (spin 2) of the perturbation, while the
symbol \textquotedblleft$\Vert$\textquotedblright\ denotes the longitudinal
part (spin 1) of the perturbation. Finally, the symbol \textquotedblleft%
$trace$\textquotedblright\ denotes the scalar part of the perturbation.
$\mathcal{N}$ is a normalization factor, $\left\langle \cdot,\cdot
\right\rangle _{x,y}$ denotes space integration and $K^{-1}$ is the inverse
\textquotedblleft\textit{propagator}\textquotedblright. We will fix our
attention to the TT tensor sector of the perturbation representing the
graviton and the scalar sector. Therefore, representation $\left(
\ref{twf}\right)  $ reduces to%
\[
\Psi\left[  h_{ij}\left(  \overrightarrow{x}\right)  \right]
\]%
\begin{equation}
=\mathcal{N}\exp\left\{  -\frac{1}{4}\left\langle hK^{-1}h\right\rangle
_{x,y}^{\bot}\right\}  \exp\left\{  -\frac{1}{4}\left\langle hK^{-1}%
h\right\rangle _{x,y}^{Trace}\right\}  . \label{tt}%
\end{equation}
Actually there is no reason to neglect longitudinal perturbations. However,
following the analysis of Mazur and Mottola of Ref.\cite{MazurMottola} on the
perturbation decomposition, we can discover that the relevant components can
be restricted to the TT modes and to the trace modes. Moreover, for certain
backgrounds TT tensors can be a source of instability as shown in
Refs.\cite{Instability}. Even the trace part can be regarded as a source of
instability. Indeed this is usually termed \textit{conformal }instability. The
appearance of an instability on the TT modes is known as non conformal
instability. This means that does not exist a gauge choice that can eliminate
negative modes. To proceed with Eq.$\left(  \ref{WDW3}\right)  $, we need to
know the action of some basic operators on $\Psi\left[  h_{ij}\right]  $. The
action of the operator $h_{ij}$ on $|\Psi\rangle=\Psi\left[  h_{ij}\right]  $
is realized by\cite{Variational}
\begin{equation}
h_{ij}\left(  x\right)  |\Psi\rangle=h_{ij}\left(  \overrightarrow{x}\right)
\Psi\left[  h_{ij}\right]  .
\end{equation}
The action of the operator $\pi_{ij}$ on $|\Psi\rangle$, in general, is%

\begin{equation}
\pi_{ij}\left(  x\right)  |\Psi\rangle=-i\frac{\delta}{\delta h_{ij}\left(
\overrightarrow{x}\right)  }\Psi\left[  h_{ij}\right]  ,
\end{equation}
while the inner product is defined by the functional integration:
\begin{equation}
\left\langle \Psi_{1}\mid\Psi_{2}\right\rangle =\int\left[  \mathcal{D}%
h_{ij}\right]  \Psi_{1}^{\ast}\left[  h_{ij}\right]  \Psi_{2}\left[
h_{kl}\right]  .
\end{equation}
We demand that
\begin{equation}
\frac{1}{V}\frac{\left\langle \Psi\left\vert \int_{\Sigma}d^{3}x\hat{\Lambda
}_{\Sigma}\right\vert \Psi\right\rangle }{\left\langle \Psi|\Psi\right\rangle
}=\frac{1}{V}\frac{\int\mathcal{D}\left[  g_{ij}\right]  \Psi^{\ast}\left[
h_{ij}\right]  \int_{\Sigma}d^{3}x\hat{\Lambda}_{\Sigma}\Psi\left[
h_{ij}\right]  }{\int\mathcal{D}\left[  g_{ij}\right]  \Psi^{\ast}\left[
h_{ij}\right]  \Psi\left[  h_{ij}\right]  } \label{vareq}%
\end{equation}
be stationary against arbitrary variations of $\Psi\left[  h_{ij}\right]  $.
Note that Eq.$\left(  \ref{vareq}\right)  $ can be considered as the
variational analog of a Sturm-Liouville problem with the cosmological constant
regarded as the associated eigenvalue. Therefore the solution of Eq.$\left(
\ref{WDW2}\right)  $ corresponds to the minimum of Eq.$\left(  \ref{vareq}%
\right)  $. The form of $\left\langle \Psi\left\vert \hat{\Lambda}_{\Sigma
}\right\vert \Psi\right\rangle $ can be computed with the help of the wave
functional $\left(  \ref{tt}\right)  $ and with the help of%
\begin{equation}
\left\{
\begin{array}
[c]{c}%
\frac{\left\langle \Psi\left\vert h_{ij}\left(  \overrightarrow{x}\right)
\right\vert \Psi\right\rangle }{\left\langle \Psi|\Psi\right\rangle }=0\\
\\
\frac{\left\langle \Psi\left\vert h_{ij}\left(  \overrightarrow{x}\right)
h_{kl}\left(  \overrightarrow{y}\right)  \right\vert \Psi\right\rangle
}{\left\langle \Psi|\Psi\right\rangle }=K_{ijkl}\left(  \overrightarrow
{x},\overrightarrow{y}\right)
\end{array}
\right.  .
\end{equation}
Since the wave functional $\left(  \ref{tt}\right)  $ separates the degrees of
freedom, we assume that%
\begin{equation}
\Lambda^{\prime}=\Lambda^{^{\prime}\bot}+\Lambda^{\prime trace},
\end{equation}
then Eq.$\left(  \ref{vareq}\right)  $ becomes%
\begin{equation}
\left\{
\begin{array}
[c]{c}%
\frac{1}{V}\frac{\left\langle \Psi\left\vert \hat{\Lambda}_{\Sigma}^{\bot
}\right\vert \Psi\right\rangle }{\left\langle \Psi|\Psi\right\rangle }%
=\Lambda^{^{\prime}\bot}\\
\\
\frac{1}{V}\frac{\left\langle \Psi\left\vert \hat{\Lambda}_{\Sigma}%
^{trace}\right\vert \Psi\right\rangle }{\left\langle \Psi|\Psi\right\rangle
}=\Lambda^{\prime trace}%
\end{array}
\right.  . \label{svareq}%
\end{equation}
Extracting the TT tensor contribution, we get%
\[
\hat{\Lambda}_{\Sigma}^{\bot}=\frac{1}{4V}\int_{\Sigma}d^{3}x\sqrt{\bar{g}%
}G^{ijkl}\left[  \left(  2\kappa\right)  K^{-1\bot}\left(  x,x\right)
_{ijkl}\right.
\]%
\begin{equation}
\left.  +\frac{1}{\left(  2\kappa\right)  }\left(  \triangle_{2}\right)
_{j}^{a}K^{\bot}\left(  x,x\right)  _{iakl}\right]  . \label{p22}%
\end{equation}
The propagator $K^{\bot}\left(  x,x\right)  _{iakl}$ can be represented as
\begin{equation}
K^{\bot}\left(  \overrightarrow{x},\overrightarrow{y}\right)  _{iakl}:=%
{\displaystyle\sum_{\tau}}
\frac{h_{ia}^{\left(  \tau\right)  \bot}\left(  \overrightarrow{x}\right)
h_{kl}^{\left(  \tau\right)  \bot}\left(  \overrightarrow{y}\right)
}{2\lambda\left(  \tau\right)  }, \label{proptt}%
\end{equation}
where $h_{ia}^{\left(  \tau\right)  \bot}\left(  \overrightarrow{x}\right)  $
are the eigenfunctions of $\triangle_{2}$. $\tau$ denotes a complete set of
indices and $\lambda\left(  \tau\right)  $ are a set of variational parameters
to be determined by the minimization of Eq.$\left(  \ref{p22}\right)  $. The
expectation value of $\hat{\Lambda}_{\Sigma}^{\bot}$ is easily obtained by
inserting the form of the propagator into Eq.$\left(  \ref{p22}\right)  $%
\[
\Lambda^{^{\prime}\bot}\left(  \lambda_{i}\right)
\]%
\begin{equation}
=\frac{1}{4}%
{\displaystyle\sum_{\tau}}
{\displaystyle\sum_{i=1}^{2}}
\left[  \left(  2\kappa\right)  \lambda_{i}\left(  \tau\right)  +\frac
{\omega_{i}^{2}\left(  \tau\right)  }{\left(  2\kappa\right)  \lambda
_{i}\left(  \tau\right)  }\right]  .
\end{equation}
By minimizing with respect to the variational function $\lambda_{i}\left(
\tau\right)  $, we obtain the total one loop energy density for TT tensors%
\begin{equation}
\Lambda^{TT}\left(  \lambda_{i}\right)  =\frac{1}{4}%
{\displaystyle\sum_{\tau}}
\left[  \sqrt{\omega_{1}^{2}\left(  \tau\right)  }+\sqrt{\omega_{2}^{2}\left(
\tau\right)  }\right]  , \label{lambda1loop}%
\end{equation}
where%
\begin{equation}
\Lambda^{TT}\left(  \lambda_{i}\right)  =\Lambda^{^{\prime}\bot}\left(
\lambda_{i}\right)  =-\Lambda^{\bot}\left(  \lambda_{i}\right)  /\kappa.
\label{lambdatt}%
\end{equation}
The above expression makes sense only for $\omega_{i}^{2}\left(  \tau\right)
>0$.

\section{The transverse traceless (TT) spin 2 operator for the Schwarzschild
metric and the W.K.B. approximation}

\label{p2}The Spin-two operator for the Schwarzschild metric is defined by%
\begin{equation}
\left(  \triangle_{2}h^{TT}\right)  _{i}^{j}:=-\left(  \triangle_{T}%
h^{TT}\right)  _{i}^{j}+2\left(  Rh^{TT}\right)  _{i}^{j}, \label{spin2}%
\end{equation}
where the transverse-traceless (TT) tensor for the quantum fluctuation is
obtained by the following decomposition%
\begin{equation}
h_{i}^{j}=h_{i}^{j}-\frac{1}{3}\delta_{i}^{j}h+\frac{1}{3}\delta_{i}%
^{j}h=\left(  h^{T}\right)  _{i}^{j}+\frac{1}{3}\delta_{i}^{j}h.
\end{equation}
This implies that $\left(  h^{T}\right)  _{i}^{j}\delta_{j}^{i}=0$. The
transversality condition is applied on $\left(  h^{T}\right)  _{i}^{j}$ and
becomes $\nabla_{j}\left(  h^{T}\right)  _{i}^{j}=0$. Thus%
\begin{equation}
-\left(  \triangle_{T}h^{TT}\right)  _{i}^{j}=-\triangle_{S}\left(
h^{TT}\right)  _{i}^{j}+\frac{6}{r^{2}}\left(  1-\frac{2MG}{r}\right)  ,
\label{tlap}%
\end{equation}
where $\triangle_{S}$ is the scalar curved Laplacian, whose form is%
\begin{equation}
\triangle_{S}=\left(  1-\frac{2MG}{r}\right)  \frac{d^{2}}{dr^{2}}+\left(
\frac{2r-3MG}{r^{2}}\right)  \frac{d}{dr}-\frac{L^{2}}{r^{2}} \label{slap}%
\end{equation}
and $R_{j\text{ }}^{a}$ is the mixed Ricci tensor whose components are:
\begin{equation}
R_{i}^{a}=\left\{  -\frac{2MG}{r^{3}},\frac{MG}{r^{3}},\frac{MG}{r^{3}%
}\right\}  ,
\end{equation}
This implies that the scalar curvature is traceless. We are therefore led to
study the following eigenvalue equation
\begin{equation}
\left(  \triangle_{2}h^{TT}\right)  _{i}^{j}=\omega^{2}h_{j}^{i} \label{p31}%
\end{equation}
where $\omega^{2}$ is the eigenvalue of the corresponding equation. In doing
so, we follow Regge and Wheeler in analyzing the equation as modes of definite
frequency, angular momentum and parity\cite{Regge Wheeler}. In particular, our
choice for the three-dimensional gravitational perturbation is represented by
its even-parity form%
\[
\left(  h^{even}\right)  _{j}^{i}\left(  r,\vartheta,\phi\right)
\]%
\begin{equation}
=diag\left[  H\left(  r\right)  ,K\left(  r\right)  ,L\left(  r\right)
\right]  Y_{lm}\left(  \vartheta,\phi\right)  , \label{pert}%
\end{equation}
with%
\begin{equation}
\left\{
\begin{array}
[c]{c}%
H\left(  r\right)  =h_{1}^{1}\left(  r\right)  -\frac{1}{3}h\left(  r\right)
\\
K\left(  r\right)  =h_{2}^{2}\left(  r\right)  -\frac{1}{3}h\left(  r\right)
\\
L\left(  r\right)  =h_{3}^{3}\left(  r\right)  -\frac{1}{3}h\left(  r\right)
\end{array}
\right.  .
\end{equation}
From the transversality condition we obtain $h_{2}^{2}\left(  r\right)
=h_{3}^{3}\left(  r\right)  $. Then $K\left(  r\right)  =L\left(  r\right)  $.
For a generic value of the angular momentum $L$, representation $\left(
\ref{pert}\right)  $ joined to Eq.$\left(  \ref{tlap}\right)  $ lead to the
following system of PDE's%

\begin{equation}
\left\{
\begin{array}
[c]{c}%
\left(  -\triangle_{S}+\frac{6}{r^{2}}\left(  1-\frac{2MG}{r}\right)
-\frac{4MG}{r^{3}}\right)  H\left(  r\right)  =\omega_{1,l}^{2}H\left(
r\right) \\
\\
\left(  -\triangle_{S}+\frac{6}{r^{2}}\left(  1-\frac{2MG}{r}\right)
+\frac{2MG}{r^{3}}\right)  K\left(  r\right)  =\omega_{2,l}^{2}K\left(
r\right)
\end{array}
\right.  . \label{p33}%
\end{equation}
Defining reduced fields%

\begin{equation}
H\left(  r\right)  =\frac{f_{1}\left(  r\right)  }{r};\qquad K\left(
r\right)  =\frac{f_{2}\left(  r\right)  }{r},
\end{equation}
and passing to the proper geodesic distance from the \textit{throat} of the
bridge%
\begin{equation}
dx=\pm\frac{dr}{\sqrt{1-\frac{2MG}{r}}}, \label{throat}%
\end{equation}
the system $\left(  \ref{p33}\right)  $ becomes%

\begin{equation}
\left\{
\begin{array}
[c]{c}%
\left[  -\frac{d^{2}}{dx^{2}}+V_{1}\left(  r\right)  \right]  f_{1}\left(
x\right)  =\omega_{1,l}^{2}f_{1}\left(  x\right) \\
\\
\left[  -\frac{d^{2}}{dx^{2}}+V_{2}\left(  r\right)  \right]  f_{2}\left(
x\right)  =\omega_{2,l}^{2}f_{2}\left(  x\right)
\end{array}
\right.  \label{p34}%
\end{equation}
with
\begin{equation}
\left\{
\begin{array}
[c]{c}%
V_{1}\left(  r\right)  =\frac{l\left(  l+1\right)  }{r^{2}}+U_{1}\left(
r\right) \\
\\
V_{2}\left(  r\right)  =\frac{l\left(  l+1\right)  }{r^{2}}+U_{2}\left(
r\right)
\end{array}
\right.  ,
\end{equation}
where we have defined $r\equiv r\left(  x\right)  $ and%
\begin{equation}
\left\{
\begin{array}
[c]{c}%
U_{1}\left(  r\right)  =\left[  \frac{6}{r^{2}}\left(  1-\frac{2MG}{r}\right)
-\frac{3MG}{r^{3}}\right] \\
\\
U_{2}\left(  r\right)  =\left[  \frac{6}{r^{2}}\left(  1-\frac{2MG}{r}\right)
+\frac{3MG}{r^{3}}\right]
\end{array}
\right.  .
\end{equation}
Note that%
\begin{equation}
\left\{
\begin{array}
[c]{c}%
U_{1}\left(  r\right)  \geq0\qquad\text{when }r\geq\frac{5MG}{2}\\
U_{1}\left(  r\right)  <0\qquad\text{when }2MG\leq r<\frac{5MG}{2}\\
\\
U_{2}\left(  r\right)  >0\text{ }\forall r\in\left[  2MG,+\infty\right)
\end{array}
\right.  . \label{negU}%
\end{equation}
The functions $U_{1}\left(  r\right)  $ and $U_{2}\left(  r\right)  $ play the
r\^{o}le of two r-dependent effective masses $m_{1}^{2}\left(  r\right)  $ and
$m_{2}^{2}\left(  r\right)  $, respectively. In order to use the WKB
approximation, we define two r-dependent radial wave numbers $k_{1}\left(
r,l,\omega_{1,nl}\right)  $ and $k_{2}\left(  r,l,\omega_{2,nl}\right)  $%
\begin{equation}
\left\{
\begin{array}
[c]{c}%
k_{1}^{2}\left(  r,l,\omega_{1,nl}\right)  =\omega_{1,nl}^{2}-\frac{l\left(
l+1\right)  }{r^{2}}-m_{1}^{2}\left(  r\right) \\
\\
k_{2}^{2}\left(  r,l,\omega_{2,nl}\right)  =\omega_{2,nl}^{2}-\frac{l\left(
l+1\right)  }{r^{2}}-m_{2}^{2}\left(  r\right)
\end{array}
\right.  \label{rwn}%
\end{equation}
for $r\geq\frac{5MG}{2}.$ When $2MG\leq r<\frac{5MG}{2}$, $k_{1}^{2}\left(
r,l,\omega_{1,nl}\right)  $ becomes%
\begin{equation}
k_{1}^{2}\left(  r,l,\omega_{1,nl}\right)  =\omega_{1,nl}^{2}-\frac{l\left(
l+1\right)  }{r^{2}}+m_{1}^{2}\left(  r\right)  . \label{rwn1}%
\end{equation}

\subsection{The trace part contribution}

\label{p3}The trace part of the perturbation can be extracted from Eq.$\left(
\ref{rexp}\right)  $ to give%
\begin{equation}
\int_{\Sigma}d^{3}x\sqrt{\bar{g}}\left[  -\frac{1}{4}h\triangle h+\frac{1}%
{2}h\nabla_{l}\nabla_{i}h^{li}\right.
\end{equation}%
\begin{equation}
\left.  +\frac{1}{2}hR_{ij}h^{ij}+\frac{1}{4}h\left(  R^{\left(  0\right)
}\right)  h\right]  . \label{tr0}%
\end{equation}
Some remark before proceeding is in order. First of all one has to note the
minus sign in front of the laplacian acting on the scalar term: this is the
origin of the conformal instability. Secondly the use of gaussian wave
functionals leads to a further simplification in Eq.$\left(  \ref{tr0}\right)
$, because mixed terms of the type $h\left(  \text{some operator}\right)
h^{TT}$ disappear. Last for some backgrounds like the one we have considered
here the three scalar curvature vanishes. Therefore the only piece of the
quadratic order coming from the scalar curvature expansion is%
\begin{equation}
\int_{\Sigma}d^{3}x\sqrt{\bar{g}}\left[  -\frac{1}{4}h\triangle h\right]  .
\end{equation}
Moreover, if we follow the orthogonal decomposition of Eq.$\left(
\ref{p21a}\right)  $, we can write%
\begin{equation}
\pi_{ij}=\frac{1}{3}\pi g_{ij}+\pi_{ij}^{long}+\pi_{ij}^{\bot},
\end{equation}
then the trace part of $\hat{\Lambda}_{\Sigma}$ becomes
\begin{equation}
-\frac{2\kappa}{6}\int_{\Sigma}d^{3}x\frac{\pi^{2}}{\sqrt{\bar{g}}}-\frac
{1}{2\kappa}\int_{\Sigma}d^{3}x\sqrt{\bar{g}}\left(  -\frac{1}{4}h\triangle
h\right)  . \label{tr1}%
\end{equation}
By repeating the scheme of calculation of section \ref{p1} to the trace part
in Eq.$\left(  \ref{tr1}\right)  $, we get the scalar part contribution to the
ZPE%
\[
\hat{\Lambda}_{\Sigma}^{trace}=-\frac{1}{4V}\int_{\Sigma}d^{3}x\sqrt{\bar{g}%
}\left[  \frac{1}{6}\left(  2\kappa\right)  K^{-1}\left(  x,x\right)  \right.
\]%
\begin{equation}
\left.  +\frac{1}{\left(  2\kappa\right)  }\left(  \triangle\right)  K\left(
x,x\right)  \right]  . \label{lambdatrace}%
\end{equation}
The propagator $K\left(  x,x\right)  $ can be represented as%
\begin{equation}
K^{\bot}\left(  \overrightarrow{x},\overrightarrow{y}\right)  :=%
{\displaystyle\sum_{\tau}}
\frac{h^{\left(  \tau\right)  }\left(  \overrightarrow{x}\right)  h^{\left(
\tau\right)  }\left(  \overrightarrow{y}\right)  }{2\lambda\left(
\tau\right)  },
\end{equation}
where $h^{\left(  \tau\right)  }\left(  \overrightarrow{x}\right)  $ are the
eigenfunctions of $\triangle$. $\tau$ denotes a complete set of indices and
$\lambda\left(  \tau\right)  $ are a set of variational parameters to be
determined by the minimization of Eq.$\left(  \ref{lambdatrace}\right)  $. The
expectation value of $\hat{\Lambda}_{\Sigma}^{trace}$ is easily obtained by
inserting the form of the propagator into Eq.$\left(  \ref{lambdatrace}%
\right)  $%
\[
\Lambda^{^{\prime}trace}\left(  \lambda_{i}\right)
\]%
\begin{equation}
=-\frac{1}{4}%
{\displaystyle\sum_{\tau}}
\left[  \frac{1}{6}\left(  2\kappa\right)  \lambda\left(  \tau\right)
+\frac{\omega^{2}\left(  \tau\right)  }{\left(  2\kappa\right)  \lambda\left(
\tau\right)  }\right]  .
\end{equation}
By minimizing with respect to the variational function $\lambda\left(
\tau\right)  $, we obtain the total one loop energy density for the scalar
component%
\begin{equation}
\Lambda^{trace}\left(  \lambda\right)  =-\frac{1}{4}\sqrt{\frac{2}{3}}%
{\displaystyle\sum_{\tau}}
\left[  \sqrt{\omega^{2}\left(  \tau\right)  }\right]  ,
\end{equation}
where $\Lambda^{trace}\left(  \lambda\right)  =\left(  \Lambda^{\prime
}\right)  ^{trace}\left(  \lambda\right)  =-\Lambda^{trace}\left(
\lambda\right)  /\kappa$. The above expression makes sense only for
$\omega^{2}\left(  \tau\right)  >0$. In the Schwarzschild background, the
operator $\triangle$ can be identified with the operator $\triangle_{S}$ of
Eq.$\left(  \ref{slap}\right)  $. If we repeat the same steps leading to
Eq.$\left(  \ref{p34}\right)  $, we get%
\begin{equation}
\left[  -\frac{d^{2}}{dx^{2}}+V\left(  r\right)  \right]  f\left(  x\right)
=\omega_{l}^{2}f\left(  x\right)  ,
\end{equation}
where we have defined%
\begin{equation}
h\left(  r\right)  =\frac{f\left(  r\right)  }{r}%
\end{equation}
and used the proper distance from the throat defined by Eq.$\left(
\ref{throat}\right)  $. In order to use a W.K.B. approximation, we define a
r-dependent radial wave number $k\left(  r,l,\omega_{nl}\right)  $%
\begin{equation}
k^{2}\left(  r,l,\omega_{nl}\right)  =\omega_{nl}^{2}-\frac{l\left(
l+1\right)  }{r^{2}}-\frac{MG}{r^{3}} \label{ktrace}%
\end{equation}

\section{One loop energy Regularization and Renormalization}

\label{p4}In this section, we proceed to evaluate Eq.$\left(
\ref{lambda1loop}\right)  $. The method is equivalent to the scattering phase
shift method and to the same method used to compute the entropy in the brick
wall model. We begin by counting the number of modes with frequency less than
$\omega_{i}$, $i=1,2$. This is given approximately by%
\begin{equation}
\tilde{g}\left(  \omega_{i}\right)  =\int\nu_{i}\left(  l,\omega_{i}\right)
\left(  2l+1\right)  , \label{p41}%
\end{equation}
where $\nu_{i}\left(  l,\omega_{i}\right)  $, $i=1,2$ is the number of nodes
in the mode with $\left(  l,\omega_{i}\right)  $, such that $\left(  r\equiv
r\left(  x\right)  \right)  $
\begin{equation}
\nu_{i}\left(  l,\omega_{i}\right)  =\frac{1}{2\pi}\int_{-\infty}^{+\infty
}dx\sqrt{k_{i}^{2}\left(  r,l,\omega_{i}\right)  }. \label{p42}%
\end{equation}
Here it is understood that the integration with respect to $x$ and $l$ is
taken over those values which satisfy $k_{i}^{2}\left(  r,l,\omega_{i}\right)
\geq0,$ $i=1,2$. With the help of Eqs.$\left(  \ref{p41},\ref{p42}\right)  $,
we obtain the one loop total energy for TT tensors%
\begin{equation}
\frac{1}{8\pi}\sum_{i=1}^{2}\int_{-\infty}^{+\infty}dx\left[  \int
_{0}^{+\infty}\omega_{i}\frac{d\tilde{g}\left(  \omega_{i}\right)  }%
{d\omega_{i}}d\omega_{i}\right]  .
\end{equation}
By extracting the energy density contributing to the cosmological constant, we
get%
\begin{equation}
\Lambda^{TT}=\frac{1}{16\pi^{2}}\int_{0}^{+\infty}\omega_{i}^{2}\sqrt
{\omega_{i}^{2}+m_{i}^{2}\left(  r\right)  }d\omega_{i}, \label{tote1loop}%
\end{equation}
where we have included an additional $4\pi$ coming from the angular
integration. We use the zeta function regularization method to compute the
energy densities $\rho_{1}$ and $\rho_{2}$. Note that this procedure is
completely equivalent to the subtraction procedure of the Casimir energy
computation where the zero point energy (ZPE) in different backgrounds with
the same asymptotic properties is involved. To this purpose, we introduce the
additional mass parameter $\mu$ in order to restore the correct dimension for
the regularized quantities. Such an arbitrary mass scale emerges unavoidably
in any regularization scheme. Then we have%
\begin{equation}
\rho_{i}\left(  \varepsilon\right)  =\frac{1}{16\pi^{2}}\mu^{2\varepsilon}%
\int_{0}^{+\infty}d\omega_{i}\frac{\omega_{i}^{2}}{\left(  \omega_{i}%
^{2}+m_{i}^{2}\left(  r\right)  \right)  ^{\varepsilon-\frac{1}{2}}},
\label{zeta}%
\end{equation}
where%
\begin{equation}
\left\{
\begin{array}
[c]{c}%
\rho_{1}=\frac{1}{16\pi^{2}}\int_{0}^{+\infty}\omega_{1}^{2}\sqrt{\omega
_{1}^{2}+m_{1}^{2}\left(  r\right)  }d\omega_{1}\\
\\
\rho_{2}=\frac{1}{16\pi^{2}}\int_{0}^{+\infty}\omega_{2}^{2}\sqrt{\omega
_{2}^{2}+m_{2}^{2}\left(  r\right)  }d\omega_{2}%
\end{array}
\right.  . \label{edens}%
\end{equation}
The integration has to be meant in the range where $\omega_{i}^{2}+m_{i}%
^{2}\left(  r\right)  \geq0$\footnote{Details of the calculation can be found
in the Appendix.}. One gets%
\begin{equation}
\rho_{i}\left(  \varepsilon\right)  =-\frac{m_{i}^{2}\left(  r\right)
}{256\pi^{2}}\left[  \frac{1}{\varepsilon}+\ln\left(  \frac{\mu^{2}}{m_{i}%
^{2}\left(  r\right)  }\right)  +2\ln2-\frac{1}{2}\right]  , \label{zeta1}%
\end{equation}
$i=1,2$. In order to renormalize the divergent ZPE, we observe that from
Eq.$\left(  \ref{tote1loop}\right)  $, after reinserting the gravitational
constant, we write%
\begin{equation}
\Lambda^{TT}=-8\pi G\left(  \rho_{1}\left(  \varepsilon\right)  +\rho
_{2}\left(  \varepsilon\right)  \right)  .
\end{equation}
To handle with the divergent energy density we extract the divergent part of
$\Lambda^{TT}$, in the limit $\varepsilon\rightarrow0$ and we set%
\begin{equation}
\Lambda^{TT,div}=\frac{G}{32\pi\varepsilon}\left(  m_{1}^{4}\left(  r\right)
+m_{2}^{4}\left(  r\right)  \right)  .
\end{equation}
Thus, the renormalization is performed via the absorption of the divergent
part into the re-definition of the bare classical constant $\Lambda^{TT}$
\begin{equation}
\Lambda^{TT}\rightarrow\Lambda_{0}^{TT}-\Lambda^{TT,div}.
\end{equation}
The remaining finite value for the cosmological constant reads%
\[
\frac{\Lambda_{0}^{TT}}{8\pi G}=\frac{1}{256\pi^{2}}\left\{  m_{1}^{4}\left(
r\right)  \left[  \ln\left(  \frac{\mu^{2}}{m_{1}^{2}\left(  r\right)
}\right)  +2\ln2-\frac{1}{2}\right]  \right.
\]%
\begin{equation}
\left.  +m_{2}^{4}\left(  r\right)  \left[  \ln\left(  \frac{\mu^{2}}%
{m_{2}^{2}\left(  r\right)  }\right)  +2\ln2-\frac{1}{2}\right]  \right\}
=\rho_{eff}^{TT}\left(  \mu,r\right)  . \label{lambda0}%
\end{equation}
The quantity in Eq.$\left(  \ref{lambda0}\right)  $ depends on the arbitrary
mass scale $\mu.$ It is appropriate to use the renormalization group equation
to eliminate such a dependence. To this aim, we impose that\cite{RGeq}%
\begin{equation}
\frac{1}{8\pi G}\mu\frac{\partial\Lambda_{0}^{TT}\left(  \mu\right)
}{\partial\mu}=\mu\frac{d}{d\mu}\rho_{eff}^{TT}\left(  \mu,r\right)  .
\label{rg}%
\end{equation}
Solving it we find that the renormalized constant $\Lambda_{0}^{TT}$ should be
treated as a running one in the sense that it varies provided that the scale
$\mu$ is changing
\begin{equation}
\Lambda_{0}^{TT}\left(  \mu,r\right)  =\Lambda_{0}^{TT}\left(  \mu
_{0},r\right)  +\frac{G}{16\pi}\left(  m_{1}^{4}\left(  r\right)  +m_{2}%
^{4}\left(  r\right)  \right)  \ln\frac{\mu}{\mu_{0}}. \label{lambdamu}%
\end{equation}
Substituting Eq.$\left(  \ref{lambdamu}\right)  $ into Eq.$\left(
\ref{lambda0}\right)  $ we find%
\begin{equation}
\frac{\Lambda_{0}^{TT}\left(  \mu_{0},r\right)  }{8\pi G}=\frac{1}{256\pi^{2}%
}\left\{  m_{1}^{4}\left(  r\right)  \left[  \ln\left(  \frac{m_{1}^{2}\left(
r\right)  }{\mu_{0}^{2}}\right)  -2\ln2+\frac{1}{2}\right]  \right.
\end{equation}%
\begin{equation}
\left.  +m_{2}^{4}\left(  r\right)  \left[  \ln\left(  \frac{m_{2}^{2}\left(
r\right)  }{\mu_{0}^{2}}\right)  -2\ln2+\frac{1}{2}\right]  \right\}  .
\end{equation}
In order to fix the dependence of $\Lambda$ on $r$ and $M$, we find the
minimum of $\Lambda_{0}^{TT}\left(  \mu_{0},r\right)  $. To this aim, last
equation can be cast into the form\footnote{Recall Eqs.$\left(  \ref{rwn}%
,\ref{rwn1}\right)  $, showing a change of sign in $m_{1}^{2}\left(  r\right)
$. Even if this is not the most appropriate notation to indicate a change of
sign in a quantity looking like a \textquotedblleft\textit{square effective
mass}\textquotedblright, this reveals useful in the zeta function
regularization and in the serch for extrema.}%
\begin{equation}
\frac{\Lambda_{0}^{TT}\left(  \mu_{0},r\right)  }{8\pi G}=\frac{\mu_{0}^{4}%
}{256\pi^{2}}\left\{  x^{2}\left(  r\right)  \left[  \ln\left(  \frac
{\left\vert x\left(  r\right)  \right\vert }{4}\right)  +\frac{1}{2}\right]
\right.
\end{equation}%
\begin{equation}
\left.  +y^{2}\left(  r\right)  \left[  \ln\left(  \frac{y\left(  r\right)
}{4}\right)  +\frac{1}{2}\right]  \right\}  , \label{renrhoeff}%
\end{equation}
where $x\left(  r\right)  =\pm m_{1}^{2}\left(  r\right)  /\mu_{0}^{2}$ and
$y\left(  r\right)  =m_{2}^{2}\left(  r\right)  /\mu_{0}^{2}$. Now we find the
extrema of $\Lambda_{0}\left(  \mu_{0};x\left(  r\right)  ,y\left(  r\right)
\right)  $ in the range $\frac{5MG}{2}\leq r$ and we get%
\begin{equation}
\left\{
\begin{array}
[c]{c}%
x\left(  r\right)  =0\\
y\left(  r\right)  =0
\end{array}
\right.  , \label{rhomin1}%
\end{equation}
which is never satisfied and%
\begin{equation}
\left\{
\begin{array}
[c]{c}%
x\left(  r\right)  =4/e\\
y\left(  r\right)  =4/e
\end{array}
\right.  \qquad\Longrightarrow\qquad\left\{
\begin{array}
[c]{c}%
m_{1}^{2}\left(  r\right)  =4\mu_{0}^{2}/e\\
m_{2}^{2}\left(  r\right)  =4\mu_{0}^{2}/e
\end{array}
\right.  . \label{rhomin2}%
\end{equation}
which implies $M=0$ and $\bar{r}=\sqrt{3e}/2\mu_{0}.$ On the other hand, in
the range $2MG\leq r<\frac{5MG}{2}$, we get again%
\begin{equation}
\left\{
\begin{array}
[c]{c}%
x\left(  r\right)  =0\\
y\left(  r\right)  =0
\end{array}
\right.  ,
\end{equation}
which has no solution and%
\begin{equation}
\left\{
\begin{array}
[c]{c}%
-m_{1}^{2}\left(  r\right)  =4\mu_{0}^{2}/e\\
m_{2}^{2}\left(  r\right)  =4\mu_{0}^{2}/e
\end{array}
\right.  , \label{rhomin3}%
\end{equation}
which implies%
\begin{equation}
\left\{
\begin{array}
[c]{c}%
\bar{M}=4\mu_{0}^{2}\bar{r}^{3}/3eG\\
\bar{r}=\sqrt{6e}/4\mu_{0}%
\end{array}
\right.  . \label{minE}%
\end{equation}
Eq.$\left(  \ref{renrhoeff}\right)  $ evaluated on the minimum, now becomes%
\begin{equation}
\Lambda_{0}^{TT}\left(  \bar{M},\bar{r}\right)  =-\frac{\mu_{0}^{4}G}%
{2e^{2}\pi}. \label{lambdamin}%
\end{equation}
It is interesting to note that thanks to the renormalization group equation
$\left(  \ref{rg}\right)  $, we can directly compute $\Lambda_{0}^{TT}$ at the
scale $\mu_{0}$ and only with the help of Eq.$\left(  \ref{lambdamu}\right)
$, we have access at the scale $\mu$.

\subsection{One loop energy Regularization and Renormalization for the trace}

\label{p5}As for the Eq.$\left(  \ref{tote1loop}\right)  $, the energy density
associated with $\Lambda^{trace}$ is%
\[
\Lambda^{trace}=-\sqrt{\frac{2}{3}}\frac{1}{64\pi^{2}}\int_{0}^{+\infty}%
\omega^{2}\sqrt{\omega^{2}-\frac{MG}{r^{3}}}d\omega
\]%
\begin{equation}
=-\sqrt{\frac{2}{3}}\frac{1}{64\pi^{2}}\int_{0}^{+\infty}\omega^{2}%
\sqrt{\omega^{2}-m^{2}\left(  r\right)  }d\omega,
\end{equation}
where we have defined%
\begin{equation}
m^{2}\left(  r\right)  =\frac{MG}{r^{3}}.
\end{equation}
The use of the zeta function regularization leads to%
\[
\rho^{trace}\left(  \varepsilon\right)  =-\sqrt{\frac{2}{3}}\frac{1}{64\pi
^{2}}\mu^{2\varepsilon}\int_{0}^{+\infty}d\omega\frac{\omega^{2}}{\left(
\omega^{2}-m^{2}\left(  r\right)  \right)  ^{\varepsilon-\frac{1}{2}}}%
\]%
\begin{equation}
=\sqrt{\frac{2}{3}}\frac{m^{4}\left(  r\right)  }{1024\pi^{2}}\left[  \frac
{1}{\varepsilon}+\ln\left(  \frac{\mu^{2}}{m^{2}\left(  r\right)  }\right)
+2\ln2-\frac{1}{2}\right]  .
\end{equation}
In order to renormalize the divergent trace contribution to ZPE, we observe
that from Eq.$\left(  \ref{tote1loop}\right)  $, we get%
\begin{equation}
\Lambda^{trace}=8\pi G\rho^{trace}.
\end{equation}
To handle with the divergent energy density we extract the divergent part of
$\Lambda^{trace}$, in the limit $\varepsilon\rightarrow0$ and we set%
\begin{equation}
\Lambda^{trace,div}=\sqrt{\frac{2}{3}}\frac{G}{128\pi\varepsilon}m^{4}\left(
r\right)  .
\end{equation}
Thus, the renormalization is performed via the absorption of the divergent
part into the re-definition of the bare classical constant $\Lambda^{trace}$
\begin{equation}
\Lambda^{trace}\rightarrow\Lambda_{0}^{trace}-\Lambda^{trace,div}.
\end{equation}
Therefore, the remaining finite value for the cosmological constant reads%
\[
\frac{\Lambda_{0}^{trace}}{8\pi G}%
\]%
\begin{equation}
=\sqrt{\frac{2}{3}}\frac{m^{4}\left(  r\right)  }{1024\pi^{2}}\left[
\ln\left(  \frac{\mu^{2}}{m^{2}\left(  r\right)  }\right)  +2\ln2-\frac{1}%
{2}\right]  =\rho_{eff}^{trace}\left(  \mu,r\right)  \label{lambda0tr}%
\end{equation}
As for the quantity in Eq.$\left(  \ref{lambda0}\right)  $, we have a
dependence on the arbitrary mass scale $\mu.$ The use of the renormalization
group equation gives%
\begin{equation}
\frac{1}{8\pi G}\mu\frac{\partial\Lambda_{0}^{trace}\left(  \mu\right)
}{\partial\mu}=\mu\frac{d}{d\mu}\rho_{eff}^{trace}\left(  \mu,r\right)  .
\label{lambda0treq}%
\end{equation}
Solving Eq.$\left(  \ref{lambda0treq}\right)  $, we find a running
$\Lambda_{0}^{trace}$ provided that the scale $\mu$ is changing
\begin{equation}
\Lambda_{0}^{trace}\left(  \mu,r\right)  =\Lambda_{0}^{trace}\left(  \mu
_{0},r\right)  +\sqrt{\frac{2}{3}}\frac{m^{4}\left(  r\right)  }{64\pi}%
G\ln\frac{\mu}{\mu_{0}}. \label{lambda0trmu}%
\end{equation}
Substituting Eq.$\left(  \ref{lambda0trmu}\right)  $ into Eq.$\left(
\ref{lambda0tr}\right)  $ we find%
\[
\frac{\Lambda_{0}^{trace}\left(  \mu_{0},r\right)  }{8\pi G}%
\]%
\begin{equation}
=-\sqrt{\frac{2}{3}}\frac{m^{4}\left(  r\right)  }{1024\pi^{2}}\left[
\ln\left(  \frac{m^{2}\left(  r\right)  }{\mu_{0}^{2}}\right)  -2\ln2+\frac
{1}{2}\right]  .
\end{equation}
If we adopt the same procedure of finding the minimum for $\Lambda_{0}%
^{trace}\left(  \mu_{0},r\right)  $ we discover that the only consistent
solution is for $M=0$. This leads to the conclusion that the trace part of the
perturbation does not contribute to the cosmological constant.

\section{Summary and Conclusions}

\label{p6}In this paper, we have considered how to extract information on the
cosmological constant using the Wheeler-De Witt equation. In particular, by
means of a variational approach and a orthogonal decomposition of the modes,
we have studied the contribution of the transverse-traceless tensors and the
trace in a Schwarzschild background. The use of the zeta function and a
renormalization group equation have led to Eq.$\left(  \ref{lambdamin}\right)
$ and recalling Eq.$\left(  \ref{lambdatt}\right)  $, we have obtained%
\begin{equation}
\left\{
\begin{array}
[c]{c}%
\Lambda_{0}^{\bot}\left(  \bar{M},\bar{r}\right)  =\frac{\mu_{0}^{4}G}%
{2e^{2}\pi}\\
\\
\Lambda_{0}^{trace}\left(  \bar{M},\bar{r}\right)  =0
\end{array}
\right.  .
\end{equation}
If we choose to fix the renormalization point $\mu_{0}=m_{p}$, we obtain
approximately $\Lambda_{0}^{\bot}\left(  \bar{M},\bar{r}\right)  \simeq
10^{37}GeV^{2}$ which, in terms of energy density is in agreement with the
estimate of Eq.$\left(  \ref{zpe}\right)  $. Once fixed the scale $\mu_{0}$,
we can see what happens at the cosmological constant at the scale $\mu$, by
means of Eq.$\left(  \ref{lambdamu}\right)  $. What we see is that the
cosmological constant is vanishing at the sub-planckian scale $\mu=m_{p}%
\exp\left(  -\frac{1}{4}\right)  $, but unfortunately is a scale which is very
far from the nowadays observations. However, the analysis is not complete.
Indeed, we have studied the spectrum in a W.K.B. approximation with the
following condition $k_{i}^{2}\left(  r,l,\omega_{i}\right)  \geq0,$ $i=1,2$.
Thus to complete the analysis, we need to consider the possible existence of
nonconformal unstable modes, like the ones discovered in
Refs.\cite{Instability}. If such an instability appears, this does not mean
that we have to reject the solution. In fact in Ref.\cite{Remo}, we have shown
how to cure such a problem. In that context, a\ model of \textquotedblleft
space-time foam\textquotedblright\ has been introduced in a large $N$ wormhole
approach reproducing a correct decreasing of the cosmological constant and
simultaneously a stabilization of the system under examination. Unfortunately
in that approach a renormalization scheme was missing and a W.K.B.
approximation on the wave function has been used to recover a
Schr\"{o}dinger-like equation. The possible next step is to repeat the scheme
we have adopted here in a large $N$ context, to recover the correct vanishing
behavior of the cosmological constant.

\appendix

\section{The zeta function regularization}

\label{app}In this appendix, we report details on computation leading to
expression $\left(  \ref{zeta}\right)  $. We begin with the following integral%
\begin{equation}
\rho\left(  \varepsilon\right)  =\left\{
\begin{array}
[c]{c}%
I_{+}=\mu^{2\varepsilon}\int_{0}^{+\infty}d\omega\frac{\omega^{2}}{\left(
\omega^{2}+m^{2}\left(  r\right)  \right)  ^{\varepsilon-\frac{1}{2}}}\\
\\
I_{-}=\mu^{2\varepsilon}\int_{0}^{+\infty}d\omega\frac{\omega^{2}}{\left(
\omega^{2}-m^{2}\left(  r\right)  \right)  ^{\varepsilon-\frac{1}{2}}}%
\end{array}
\right.  , \label{rho}%
\end{equation}
with $m^{2}\left(  r\right)  >0$.

\subsection{$I_{+}$ computation}

\label{app1}If we define $t=\omega/\sqrt{m^{2}\left(  r\right)  }$, the
integral $I_{+}$ in Eq.$\left(  \ref{rho}\right)  $ becomes%
\[
\rho\left(  \varepsilon\right)  =\mu^{2\varepsilon}m^{4-2\varepsilon}\left(
r\right)  \int_{0}^{+\infty}dt\frac{t^{2}}{\left(  t^{2}+1\right)
^{\varepsilon-\frac{1}{2}}}%
\]%
\[
=\frac{1}{2}\mu^{2\varepsilon}m^{4-2\varepsilon}\left(  r\right)  B\left(
\frac{3}{2},\varepsilon-2\right)
\]%
\[
\frac{1}{2}\mu^{2\varepsilon}m^{4-2\varepsilon}\left(  r\right)  \frac
{\Gamma\left(  \frac{3}{2}\right)  \Gamma\left(  \varepsilon-2\right)
}{\Gamma\left(  \varepsilon-\frac{1}{2}\right)  }%
\]%
\begin{equation}
=\frac{\sqrt{\pi}}{4}m^{4}\left(  r\right)  \left(  \frac{\mu^{2}}%
{m^{2}\left(  r\right)  }\right)  ^{\varepsilon}\frac{\Gamma\left(
\varepsilon-2\right)  }{\Gamma\left(  \varepsilon-\frac{1}{2}\right)  },
\end{equation}
where we have used the following identities involving the beta function%
\begin{equation}
B\left(  x,y\right)  =2\int_{0}^{+\infty}dt\frac{t^{2x-1}}{\left(
t^{2}+1\right)  ^{x+y}}\qquad\operatorname{Re}x>0,\operatorname{Re}y>0
\end{equation}
related to the gamma function by means of%
\begin{equation}
B\left(  x,y\right)  =\frac{\Gamma\left(  x\right)  \Gamma\left(  y\right)
}{\Gamma\left(  x+y\right)  }.
\end{equation}
Taking into account the following relations for the $\Gamma$-function%
\begin{equation}%
\begin{array}
[c]{c}%
\Gamma\left(  \varepsilon-2\right)  =\frac{\Gamma\left(  1+\varepsilon\right)
}{\varepsilon\left(  \varepsilon-1\right)  \left(  \varepsilon-2\right)  }\\
\\
\Gamma\left(  \varepsilon-\frac{1}{2}\right)  =\frac{\Gamma\left(
\varepsilon+\frac{1}{2}\right)  }{\varepsilon-\frac{1}{2}}%
\end{array}
, \label{gamma}%
\end{equation}
and the expansion for small $\varepsilon$%
\begin{equation}%
\begin{array}
[c]{cc}%
\Gamma\left(  1+\varepsilon\right)  = & 1-\gamma\varepsilon+O\left(
\varepsilon^{2}\right) \\
& \\
\Gamma\left(  \varepsilon+\frac{1}{2}\right)  = & \Gamma\left(  \frac{1}%
{2}\right)  -\varepsilon\Gamma\left(  \frac{1}{2}\right)  \left(  \gamma
+2\ln2\right)  +O\left(  \varepsilon^{2}\right) \\
& \\
x^{\varepsilon}= & 1+\varepsilon\ln x+O\left(  \varepsilon^{2}\right)
\end{array}
,\qquad
\end{equation}
where $\gamma$ is the Euler's constant, we find%
\begin{equation}
\rho\left(  \varepsilon\right)  =-\frac{m^{4}\left(  r\right)  }{16}\left[
\frac{1}{\varepsilon}+\ln\left(  \frac{\mu^{2}}{m^{2}\left(  r\right)
}\right)  +2\ln2-\frac{1}{2}\right]  .
\end{equation}

\subsection{$I_{-}$ computation}

\label{app2}If we define $t=\omega/\sqrt{m^{2}\left(  r\right)  }$, the
integral $I_{-}$ in Eq.$\left(  \ref{rho}\right)  $ becomes%
\[
\rho\left(  \varepsilon\right)  =\mu^{2\varepsilon}m^{4-2\varepsilon}\left(
r\right)  \int_{0}^{+\infty}dt\frac{t^{2}}{\left(  t^{2}-1\right)
^{\varepsilon-\frac{1}{2}}}%
\]%
\[
=\frac{1}{2}\mu^{2\varepsilon}m^{4-2\varepsilon}\left(  r\right)  B\left(
\varepsilon-2,\frac{3}{2}-\varepsilon\right)
\]%
\[
\frac{1}{2}\mu^{2\varepsilon}m^{4-2\varepsilon}\left(  r\right)  \frac
{\Gamma\left(  \frac{3}{2}-\varepsilon\right)  \Gamma\left(  \varepsilon
-2\right)  }{\Gamma\left(  -\frac{1}{2}\right)  }%
\]%
\begin{equation}
=-\frac{1}{4\sqrt{\pi}}m^{4}\left(  r\right)  \left(  \frac{\mu^{2}}%
{m^{2}\left(  r\right)  }\right)  ^{\varepsilon}\Gamma\left(  \frac{3}%
{2}-\varepsilon\right)  \Gamma\left(  \varepsilon-2\right)  ,
\end{equation}
where we have used the following identity involving the beta function%
\begin{equation}%
\begin{array}
[c]{c}%
\frac{1}{p}B\left(  1-\nu-\frac{\mu}{p},\nu\right)  =\int_{1}^{+\infty
}dtt^{\mu-1}\left(  t^{p}-1\right)  ^{\nu-1}\\
\\
p>0,\operatorname{Re}\nu>0,\operatorname{Re}\mu<p-p\operatorname{Re}\nu
\end{array}
\end{equation}
and the reflection formula%
\begin{equation}
\Gamma\left(  z\right)  \Gamma\left(  1-z\right)  =-z\Gamma\left(  -z\right)
\Gamma\left(  z\right)
\end{equation}
From the first of Eqs.$\left(  \ref{gamma}\right)  $ and from the expansion
for small $\varepsilon$%
\[
\Gamma\left(  \frac{3}{2}-\varepsilon\right)  =\Gamma\left(  \frac{3}%
{2}\right)  \left(  1-\varepsilon\left(  -\gamma-2\ln2+2\right)  \right)
+O\left(  \varepsilon^{2}\right)
\]%
\begin{equation}
x^{\varepsilon}=1+\varepsilon\ln x+O\left(  \varepsilon^{2}\right)  ,
\end{equation}
we find%
\begin{equation}
\rho\left(  \varepsilon\right)  =-\frac{m^{4}\left(  r\right)  }{16}\left[
\frac{1}{\varepsilon}+\ln\left(  \frac{\mu^{2}}{m^{2}\left(  r\right)
}\right)  +2\ln2-\frac{1}{2}\right]  .
\end{equation}


\begin{thebibliography}{99}                                                                                               %


\bibitem {Lambda}For a pioneering review on this problem see S. Weinberg,
\textsl{Rev. Mod. Phys. }\textbf{61}, 1 (1989). For more recent and detailed
reviews see V. Sahni and A. Starobinsky, \textsl{Int. J. Mod. Phys.}
\textbf{D} \textbf{9}, 373 (2000), astro-ph/9904398; N. Straumann, \textit{The
history of the cosmological constant problem} gr-qc/0208027; T.Padmanabhan,
\textsl{Phys.Rept.} \textbf{380}, 235 (2003), hep-th/0212290.

\bibitem {De Witt}B. S. DeWitt, \textsl{Phys. Rev.} \textbf{160}, 1113 (1967).

\bibitem {BergerEbin}M. Berger and D. Ebin, \textsl{J. Diff. Geom.}
\textbf{3}, 379 (1969).

\bibitem {York}J. W. York Jr., \textsl{J. Math. Phys.}, \textbf{14}, 4 (1973);
\textsl{Ann. Inst. Henri Poincar\'{e}} \textbf{A} \textbf{21}, 319 (1974).

\bibitem {MazurMottola}P. O. Mazur and E. Mottola, \textsl{Nucl. Phys.}
\textbf{B 341}, 187 (1990).

\bibitem {Vassilevich}D.V. Vassilevich, \textsl{Int. J. Mod. Phys. }\textbf{A}
\textbf{8}, 1637 (1993). D. V. Vassilevich, \textsl{Phys. Rev.} \textbf{D 52}
999 (1995); gr-qc/9411036.

\bibitem {Instability}D.J. Gross, M.J. Perry and L.G. Yaffe, \textsl{Phys.
Rev.} \textbf{D} \textbf{25}, (1982) 330. B. Allen, \textsl{Phys. Rev.}
\textbf{D 30} 1153 (1984). E. Witten, \textsl{Nucl Phys} \textbf{B 195} 481
(1982). P. Ginsparg and M.J. Perry, \textsl{Nucl. Phys.} \textbf{B}
\textbf{222} (1983) 245. R.E. Young, \textsl{Phys. Rev.} \textbf{D}
\textbf{28, (}1983) 2436. R.E. Young, \textsl{Phys. Rev.} \textbf{D}
\textbf{28} (1983) 2420. S.W. Hawking and D.N. Page, \textsl{Commun. Math.
Phys.} \textbf{87}, 577 (1983). R. Gregory and R. Laflamme, \textsl{Phys.
Rev.} \textbf{D 37} 305 (1988). R. Garattini, \textsl{Int. J. Mod. Phys.
}\textbf{A} \textbf{14}, 2905 (1999); gr-qc/9805096. E Elizalde, S Nojiri and
S D Odintsov, \textsl{Phys. Rev.} \textbf{D} \textbf{59} (1999) 061501; hep-th
9901026. M.S. Volkov and A. Wipf, \textsl{Nucl. Phys.} \textbf{B 582} (2000),
313; hep-th/0003081. R. Garattini, \textsl{Class. Quant. Grav.} \textbf{17},
3335 (2000); gr-qc/0006076. T. Prestidge, \textsl{Phys. Rev.} \textbf{D}
\textbf{61}, (2000) 084002; hep-th/9907163. S.S. Gubser and I. Mitra,
\textit{Instability of charged black holes in Anti-de Sitter space},
hep-th/0009126; R. Garattini, \textsl{Class. Quant. Grav.} \textbf{18}, 571
(2001); gr-qc/0012078. S.S. Gubser and I. Mitra, \textsl{JHEP} \textbf{8}
(2001) 18. J.P. Gregory and S.F. Ross, \textsl{Phys. Rev.} \textbf{D}
\textbf{64} (2001) 124006, hep-th/0106220. H.S. Reall, \textsl{Phys. Rev.}
\textbf{D} \textbf{64} (2001) 044005, hep-th/0104071. G. Gibbons and S.A.
Hartnoll, \textsl{Phys. Rev.} \textbf{D} \textbf{66} (2001) 064024, hep-th/0206202.

\bibitem {Variational}A. K. Kerman and D. Vautherin, \textsl{Ann. Phys.},
\textbf{192}, 408 (1989); J. M. Cornwall, R. Jackiw and E. Tomboulis,
\textsl{Phys. Rev.} \textbf{D 8}, (1974) 2428; R. Jackiw in
\textit{S\'{e}minaire de Math\'{e}matiques Sup\'{e}rieures, Montr\'{e}al,
Qu\'{e}bec, Canada- June 1988 - Notes by P. de Sousa Gerbert}; M. Consoli and
G. Preparata, \textsl{Phys. Lett.} \textbf{B}, \textbf{154}, 411 (1985).

\bibitem {Regge Wheeler}T. Regge and J. A. Wheeler, \textsl{Phys. Rev.
}\textbf{108}, 1063 (1957).

\bibitem {RGeq}J.Perez-Mercader and S.D. Odintsov, \textsl{Int. J. Mod. Phys.}
\textbf{D} \textbf{1}, 401 (1992). I.O. Cherednikov, \textsl{Acta Physica
Slovaca}, \textbf{52}, (2002), 221. I.O. Cherednikov, \textsl{Acta Phys.
Polon.} \textbf{B} \textbf{35}, 1607 (2004). M. Bordag, U. Mohideen and V.M.
Mostepanenko, \textsl{Phys. Rep.} \textbf{353}, 1 (2001). Inclusion of
non-perturbative effects, namely beyond one-loop, in de Sitter Quantum Gravity
have been discussed in S. Falkenberg and S. D. Odintsov, \textsl{Int. J. Mod.
Phys. }\textbf{A} \textbf{13}, 607 (1998); hep-th 9612019.

\bibitem {Remo}R. Garattini, \textsl{Int. J. Mod. Phys. }\textbf{D}
\textbf{4}, 635 (2002); gr-qc/0003090.

\bibitem {GR}I.S. Gradshteyn and I.M. Ryzhik, \textit{Table of Integrals,
Series, and Products }(corrected and enlarged edition), edited by A. Jeffrey
(Academic Press, Inc.).
\end{thebibliography}
\end{document}